\documentstyle[12pt]{article}

\newlength{\dinwidth}
\newlength{\dinmargin}
\newcommand{\resection}[1]{\setcounter{equation}{0}\section{#1}}

\begin{document}
\vspace*{4cm}
\begin{center}
  \begin{Large}
  \begin{bf}
On the coupling of heavy mesons to pions in QCD\\
  \end{bf}
  \end{Large}
  \vspace{10mm}
  \begin{large}
P. Colangelo and G. Nardulli\\
  \end{large}
Dipartimento di Fisica, Univ.di Bari\\
I.N.F.N., Sezione di Bari\\
  \vspace{5mm}
  \begin{large}
A. Deandrea, N. Di Bartolomeo and R. Gatto\\
  \end{large}
D\'epartement de Physique Th\'eorique, Univ. de Gen\`eve\\
  \vspace{5mm}
  \begin{large}
F. Feruglio\\
  \end{large}
Dipartimento di Fisica, Univ. di Padova\\
  \vspace{5mm}
\end{center}
  \vspace{2cm}
\begin{center}
UGVA-DPT 1994/06-856\\\
BARI-TH/94-171 \\\
hep-ph/9406295\\\
June 1994
\end{center}
\vspace{1cm}
\noindent
$^*$ Partially supported by the Swiss National  Science Foundation
\newpage
\thispagestyle{empty}
\begin{quotation}
\vspace*{5cm}
\begin{center}
  \begin{bf}
  ABSTRACT

   \end{bf}
\end{center}
 \vspace*{1cm}
We compute the strong coupling constant $g_{P^* P \pi}$, with $P$ and $P^*$
respectively pseudoscalar and vector heavy mesons by using the QCD sum rules
approach. Our computation is based on the evaluation of the time ordered
product of currents between the vacuum and the soft pion state.
The so-called parasitic terms are taken into account and give a contribution to
the sum rule of the same order of the lowest lying state, while
higher dimension non perturbative terms have small numerical effects.
The infinite heavy quark mass limit is also examined.
  \vspace{5mm}
\noindent

\end{quotation}
\newpage
\setcounter{page}{1}
\def\lq{\left [}
\def\rq{\right ]}
\def\qq{<{\overline u}u>}
\def\dmu{\partial_{\mu}}
\def\dmus{\partial^{\mu}}
\def\gi{{g_{P^* P \pi}}}

\def\gid{{g_{D^* D \pi}}}
\def\gib{{g_{B^* B \pi}}}
\def\slash#1{\setbox0=\hbox{$#1$}#1\hskip-\wd0\dimen0=5pt\advance
       \dimen0 by-\ht0\advance\dimen0 by\dp0\lower0.5\dimen0\hbox
	 to\wd0{\hss\sl/\/\hss}}
\newcommand{\be}{\begin{equation}}
\newcommand{\ee}{\end{equation}}
\newcommand{\bea}{\begin{eqnarray}}
\newcommand{\eea}{\end{eqnarray}}
\newcommand{\nn}{\nonumber}
\newcommand{\dd}{\displaystyle}

\resection{Introduction}
In this letter we shall address the problem of the determination of the strong
coupling constant $\gi$, where $P$ and $P^*$ are respectively the lowest
pseudoscalar and
vector heavy meson made up of a light $q$ antiquark and a heavy quark $Q$.
 This
coupling is defined by the strong amplitude
\be
<\pi^-(q)~P^o(q_2) | P^{*-}(q_1,\epsilon )> \; = \; \gi \;
\epsilon^{\mu} \cdot q_{\mu}
\label{sda}
\ee
and is of physical interest for a number of reasons. First, for $P=D$ and
$P^*=D^*$, the decay amplitude (\ref{sda}) describes a process which is
observed
experimentally. Present data provide the upper bound:
\be
\gid \le 13.8
\ee
 from $\Gamma_{tot}(D^{*+}) < 131$ KeV \cite{acc} and
$BR(D^{*+} \to D^o \pi^+) = 30.8 \pm 0.4 \pm 0.8$ \cite {cleo}.
Second, it is commonly believed that the form
factor $F_1 (q^2)$ describing the semileptonic decay $B \to \pi \ell \nu$ is
dominated, for $q^2 \simeq q^2_{max}=(m_B-m_{\pi})^2$, by the $B^*$ pole; in
this case $F_1 (q^2_{max})$ would be given by the formula \cite{nuss,wolf}:
\be
F_1 (q^2_{max}) \simeq {{f_{B^*}}\over{4}}
{{g_{B^* B \pi}}\over{m_{\pi}+(m_{B^*}-m_B)}}
\ee
so that knowledge of the strong coupling $g_{B^* B \pi}$ would shed light on
this decay process ($f_{B^*}$ is the leptonic decay constant of $B^*$).
Finally in the
chiral effective field theories for heavy mesons that have recently received
attention by a number of authors \cite{wise}, \cite{yan}, \cite{burdman},
\cite{casalbuoni}, the above mentioned coupling is related to the interaction
lagrangian
\footnote{\noindent We employ the notations of \cite{wise}, where
$H=(1+{\slash v})/2 (P^*_{\mu} \gamma^{\mu} -P_5 \gamma_5)$, ($v$ is the
heavy meson velocity, $P^*_{\mu}$ and $P_5$ are annihilation operators of the
heavy mesons $P^*$ and $P$ respectively) and $\xi^2=\Sigma=\exp i M/f$, with
$M$ the $3 \times 3$ matrix of the Nambu-Goldstone pseudoscalar meson octet and
$f=f_{\pi}$ in the chiral limit
 ($f_{\pi}=132$ MeV).}
\be
{\cal L}_{int}= {i\over 2} \; g \;
Tr [H\gamma_{\mu} \gamma_5 (\xi^{\dag} \dmu \xi -
\xi \dmu \xi^{\dag}) {\overline H}]
\label{lint}
\ee
by the formula
\be
\gi={{2 m_P}\over{f}} g \; . \label{ginf}
\ee
The coupling constant $g$ appears in a number of calculations that make use of
the chiral effective theories in the infinite heavy quark mass limit
\cite{casalbuoni}, \cite{georgi}, \cite{amundson}, \cite{defazio}.

The computation we shall describe in the next sections is based on the QCD sum
rules approach \footnote{For a review of QCD sum rules as applied to heavy
masses see \cite{paver}.}. Some other QCD sum rules calculations of the strong
coupling constant $\gi$ can be found in the literature \cite{eletski},
\cite{grozin}.
The authors of ref.\cite{eletski} compute $g_{D^* D \pi}$ by considering
two-point functions in an external pion field, whereas the authors of
ref.\cite{grozin} compute the coupling constant $g$ of eq.(\ref{lint}).
We differ
 from these calculations in several aspects. First of all we compute
$\gid$, $\gib$ and $g$, which allows us to discuss
$1/{m_Q}$ corrections to the asymptotic $m_Q \to \infty$ limit. Second we
use the method of ref.\cite{novikov} both in the evaluation of the correlators
(the time ordered product of currents is taken between the vacuum and the pion
state) and in the treatment of the so called parasitic terms, arising from
non-diagonal transitions \footnote{See also \cite{smilga}}; they are present
if, as in our case, one takes the soft pion limit $q \to 0$, which implies
$q_1=q_2$ and one has to take the single and not the
double Borel transform. These
differences alter the numerical result, as we shall discuss below; other
differences with refs.\cite{eletski}, \cite{grozin}, arising from higher order
non perturbative terms that we have included,
have small numerical effects in the final results.

\resection{QCD sum rule calculation of $g_{D^* D \pi}$ and $g_{B^* B \pi}$}

For definiteness we shall consider the off-shell process
\be
B^{*-}(\epsilon,q_1) \to {\bar B^0} (q_2) + \pi^-(q) \; ,
\ee
the case of $D^* \to D \pi$ being completely analogous. We consider the
correlator
\be
A_{\mu}(P,q) = i \int dx <\pi^-(q)| T(V_{\mu}(x) j_5(0) |0> e^{-iq_1x} = A
q_{\mu} + B P_{\mu}
\label{corr}
\ee
where $V_{\mu}={\overline u} \gamma_{\mu} b$, $j_5=i{\overline b} \gamma_5 d$,
$P=q_1+q_2$ and $A$, $B$ are scalar functions of $q_1^2$, $q_2^2$, $q^2$.

Both $A$ and $B$ satisfy dispersion relations and are computed, according to
the QCD sum rules method, in two ways: either by saturating the dispersion
relation by physical hadronic states or by means of the operator product
expansion (OPE). We shall focus our attention on the invariant function $A$
that we compute in the soft pion limit ($q \to 0$) and for large Euclidean
momenta ($q_1^2=q_2^2 \to - \infty$).

Performing the OPE on $A$ we get the following result
\be
A=A^{(0)}+A^{(1)}+A^{(2)}+A^{(3)}+A^{(4)}+A^{(5)}
\label{terms}
\ee
with
\bea
A^{(0)}&=&{{-1}\over{q_1^2-m_b^2}} \lq m_b f_{\pi} +{{\qq}\over{f_{\pi}}}\rq
\nn\\
A^{(1)}&=&-{2\over 3} {{1}\over{q_1^2-m_b^2}} {{\qq}\over{f_{\pi}}} \lq
{{m_b^2}\over{q_1^2-m_b^2}} -2 \rq \nn\\
A^{(2)}&=&{{m_b f_{\pi} m_1^2}\over{9 (q_1^2-m_b^2)^2}} \lq 1+ {{10
m_b^2}\over{q_1^2- m_b^2}} \rq - {{m_o^2 \qq}\over{4 f_{\pi} (q_1^2-m_b^2)^2}}
\lq 1- {{2 m_b^2}\over{q_1^2 m_b^2}} \rq \nn\\
A^{(3)}&=&{{m_0^2 \qq}\over{6 f_{\pi}}} \lq {1\over{(q_1^2-m_b^2)^2}} -{{2
m_b^2}\over{(q_1^2-m_b^2)^3}} +{{6 m_b^4}\over{(q_1^2-m_b^2)^4}} \rq \nn\\
A^{(4)}&=&{1\over{(q_1^2-m_b^2)^2}} \lq {{m_0^2 \qq}\over{4 f_{\pi}}} +m_b
f_{\pi} m_1^2 \rq \nn\\
A^{(5)}&=&{{m_0^2 \qq}\over{6 f_{\pi}}} \lq {1\over{(q_1^2-m_b^2)^2}} -{{2
m_b^2}\over{(q_1^2-m_b^2)^3}} \rq \; .
\label{cntr}
\eea
In eqs.(\ref{cntr}) $\qq$ is the quark condensate ($\qq =-(240 MeV)^3$), $m_0$
and $m_1$ are defined by the equations
\be
<{\overline u} g_s \sigma \cdot G u> =m_0^2\qq
\ee
and
\be
<\pi (q)|{\overline u} D^2 \gamma_{\mu} \gamma_5 d |0> = -i f_{\pi} m_1^2
q_{\mu}
\ee
and their numerical values are: $m_0^2=0.8 \; GeV^2$, $m_1^2=0.2 \; GeV^2$
\cite{novikov,chernyak}.
The origin of the different terms in (\ref{terms}) is as follows.
$A^{(0)}$ is the leading term in the short distance expansion; $A^{(1)}$,
$A^{(2)}$ and $A^{(3)}$ arise from the expansion of $V_{\mu}(x)$ at the
first, second and third order in powers of $x$; $A^{(4)}$ and $A^{(5)}$ arise
 from the expansion of the heavy quark propagator at the second order and from
the zeroth and first term in the expansion of $V_{\mu}(x)$ respectively. We
stress that we have considered all the operators with dimension  $D\le 5$ in
the OPE of the currents appearing in (\ref{corr});
we also note that, with the obvious change $b \to c$, eq.(\ref{cntr})
also applies to the amplitude for charm decay $D^* \to D \pi$.

Let us now evaluate the hadronic side of the sum rule, which can be obtained by
writing for $A(0,q_1^2,q_2^2)$ the dispersion relation
\be
A(0,q_1^2,q_2^2)={1\over{\pi^2}} \int ds ds' {{\rho(s,s')}\over{(s-q_1^2)
(s'-q_2^2)}} + subtractions \; .
\label{intg}
\ee
We divide the integration region into three parts. The first region (I) is the
square given by $ m_b^2 \le s \le s_0$, $m_b^2 \le s'\le s_0$;
for $s_0$ small enough, (I) contains
only the $B$ and $B^*$ poles, whose contribution is
\bea
A_I(0,q_1^2,q_2^2)&=&{{f_B f_{B^*} m_B^2}\over{4 m_b m_{B^*}}} \;
\Big[ {{\gib (3
m_{B^*}^2+m_B^2)}\over{(q_1^2-m_{B^*}^2) (q_2^2-m_B^2)}} +\nn\\
&+& {{\gib}\over{q_1^2-m_{B^*}^2}}+{{3 f_+ - f_-}\over{q_2^2 -m_B^2}} \Big]
\label{uno}
\eea
where $f_B$, $f_{B^*}$, $f_+$ and $f_-$ are defined by
\be
<0|A_{\mu}(0)|B(p)> = i p_{\mu} f_B
\ee
\be
<0|V_{\mu}(0)|B^*(\epsilon,p)>= \epsilon_{\mu} f_{B^*} m_{B^*}
\ee
\be
<\pi^-(q) {\overline B}^0(q_2) | B^{*-}(q_1)> = (f_+ P_{\mu} - f_- q_{\mu})
\epsilon^{\mu} \; .
\ee
Similar definitions hold for the $D^* \to D \pi$ decay.

The second (II) integration region in (\ref{intg}) is defined as follows:
$m_b^2 \le s \le s_0$ and $s'>s_0$ or $m_b^2 \le s' \le s_0$ and $s>s_0$.
In this region we
can distinguish two contributions. The first one is obtained by coupling the
vector current $V_{\mu}$ to the pion and the $B$ and the second one arises by
coupling the pseudoscalar $j_5$ current to $\pi$ and $B^*$. The first term is
as follows
\be
A_{II}(0,q_1^2,q_2^2)={{f_B m_B^2}\over{2 m_b}} {1\over{q_2^2 -m_B^2}} \lq 3F_1
-{{q_2^2}\over{q_1^2}} (F_0-F_1)\rq
\ee
where $F_0$ and $F_1$ are the usual form factors \cite{stech} describing the
coupling of the vector current $V_{\mu}$ to the $B$ and $\pi$. We can safely
assume that $F_1=F_1(q_1^2)$ is dominated by the $B^*$ pole \cite{ball},
\cite{santorelli}, \cite{grinstein}, therefore the contribution from this form
factor would be omitted to avoid double counting; assuming that the $q_1^2$
dependence of $F_0$ is given by the low lying $0^+$ pole (whose mass we denote
by $\overline m$) we obtain
\be
A_{II}(0,q_1^2,q_2^2)={{f_B m_B^2 {\overline m}^2 F_0(0)}\over{2 m_b}}
{1\over{q_2^2 -m_B^2}}{1\over{q_1^2 -{\overline m}^2}}
\label{due}
\ee
As for the second term in the II region, it would be proportional to the
$A_0(q_2^2)$ form factor (in the notations of ref.\cite{stech}). Assuming again
the dominance of the low lying $0^-$ $B$-pole we would obtain a term which
coincides with $A_I$ and therefore it will be omitted to avoid double counting.

Let us finally consider the third (III) region in the ($s,s'$) plane, as
defined by $s,s'>s_0$. Assuming duality, in
this region $A_{III}$ should coincide with the asymptotic ($q_1^2=q_2^2 \to
-\infty$) limit of (\ref{terms}); therefore the spectral function is given by:
\be
\rho_{III}(s,s')= \lq m_b f_{\pi} -{{\qq}\over{3 f_{\pi}}} \rq \pi^2 \theta
(s-s_0) \theta (s'-s_0) \delta(s-s') \; .
\label{tre}
\ee
Let us now discuss, before taking the single Borel transform of
$A(0,q_1^2,q_2^2)$, the role of the terms (\ref{due}), (\ref{tre})
 and the single pole contributions to $A_I$ in (\ref{uno})
 (in the literature they are called
``parasitic" or non-diagonal terms). In general these terms contain new unknown
quantities as well as several uncertainties; however we can exploit their
different dependence on the Borel parameter $M^2$ as compared
to the resonance term (the first contribution in (\ref{uno}) ); as a matter of
fact, putting $q_1^2=q_2^2$ and taking the Borel transform, one gets the
following sum rule:
\bea
{\gamma \over {M^2}}+{{\lambda(M^2)}} &=& \exp(\Omega/M^2) \;
\Big(  \lq m_b f_{\pi} - {{\qq}\over{3 f_{\pi}}} \rq \nn\\
&+&{1\over{M^2}} \lq -{{2\qq m_b^2}\over{3 f_{\pi}}}+{{10 m_b f_{\pi} m_1^2}
\over{9}} + {{m_0^2\qq}\over{3 f_{\pi}}} \rq \nn\\
&+& {1\over{2 M^4}} \lq -{{10 m_b^3 f_{\pi} m_1^2}\over{9}}+{{m_0^2\qq m_b^2}
\over{6 f_{\pi}}} \rq +\nn\\
&+& {{m_0^2\qq m_b^4}\over{6 f_{\pi} M^6}} \Big)
\label{sr}
\eea
where
\be
\gamma={{f_B ~f_{B^*} ~m_B^2 ~\gib}\over{4 m_b m_{B^*}}} \;
 (3 m^2_{B^*} + m^2_B)
\ee
%\be
%\gamma= f_B ~f_{B^*} ~m_B^2 ~\gib
%\ee

and $\Omega = (m_B^2+m_{B^*}^2-2m_b^2)/2$, while $\lambda (M^2)$ is given by
\be
\lambda (M^2)=d_0+d_1 e^{-\delta/M^2}
\label{lam}
\ee
with
\be
\delta = s_0 -m_B^2
\label{del}
\ee
and $d_0$, $d_1$ unknown quantities. We note that
$\lambda (M^2)$ takes into account all the possible parasitic terms, either
deriving from eqs.(\ref{due}),(\ref{tre})
or from the last two terms on the r.h.s.
of (\ref{uno}). We also note that in deriving (\ref{lam}) and (\ref{del})
 we put
${\overline m}^2 \simeq s_0$, which, for the usually accepted values of $s_0$
($s_0\simeq 33 - 36 \; GeV^2$ for the $B$, $s_0 \simeq 6 - 8 \; GeV^2$ for the
charm
case) is a reasonable approximation. Moreover we put $m_B^2 \simeq
(m_B^2+m_{B^*}^2)/2$ and $m_D^2 \simeq (m_D^2+m_{D^*}^2)/2$.

We can now differentiate (\ref{sr})
in the variable $1/M^2$ so that we remain with a new
sum rule (NSR) for the variable
$\gamma +\partial \lambda/\partial (1/M^2)$. Since $\gamma$ does not depend on
$M^2$, we differentiate again this equation, so that we obtain an equation in
the unknown quantity $d_1$. Substituting this expression in  NSR we finally
obtain the following sum rule for $\gib$:
\bea
\gamma &=& \exp \lq {\Omega \over {M^2}}
\rq \Big \{ \Omega \left( 1+ {\Omega \over \delta} \right) \Big[ m_b f_{\pi} -
{\qq \over {3 f_{\pi}}} \left( 1+{{2 m_b^2}\over{M^2}} \right) +\nn\\
&+& {{10 f_{\pi} m_1^2 m_b}\over{9 M^2}} \left( 1-{{m_b^2}\over{2 M^2}} \right)
+ {{m_0^2 \qq}\over{3 f_{\pi} M^2}} \left( 1+{{m_b^2}\over{4 M^2}} + {{m_b^4}
\over{2 M^4}} \right) \Big] +\nn\\
&+&{{2\Omega +\delta}\over{\delta}} \Big[ -{{2\qq m_b^2}\over{3 f_{\pi}}} +
{{10 f_{\pi} m_1^2 m_b}\over{9}} \left(1-{{m_b^2}\over{M^2}} \right) + \nn\\
&+& {{m_0^2\qq}\over{6 f_{\pi}}} \left( 2+ {{m_b^2}\over {M^2}} + {{3 m_b^4}
\over{M^4}} \right) \Big] -{{10 f_{\pi} m_1^2 m_b^3}\over{9 \delta}} + \nn\\
&+&{{m_0^2 \qq
m_b^2}\over{6 f_{\pi} \delta}} \left( 1+{{6 m_b^2}\over{M^2}} \right)\Big\}
\label{sr1}
\eea
The analysis of the sum rule follows the usual criteria
adopted in the literature: we check the existence of a region in $M^2$ where
the OPE displays a convergent structure and where the highest states give a
small contribution, as we discuss in the following section.
Using
$m_b=4.6 \; GeV$ and $m_c= 1.34 \; GeV$ \cite{paver} for the heavy
quark masses we obtain, in the region $M^2 \ge 30 \; GeV^2$ for $B$
and $M^2 \ge 8 \; GeV^2$ for $D$:
\bea f_B \; f_{B^*} \; \gib & = &  0.56 \pm 0.10 \; GeV^2  \label{res1}\\
 f_D \; f_{D^*}\; \gid & = &  0.34 \pm 0.04 \; GeV^2  \; ; \label{res2} \eea
\noindent
the uncertainty is mainly due to the variation of the threshold $s_0$.
It should be noticed that the parasitic terms represent a relevant fraction
($40 - 50 \%$) of the sum rule.

The strong couplings $\gi$ can be derived from
(\ref{res1}, \ref{res2}) once the leptonic constants are known.
Some authors \cite{eletski,grozin} suggest that in eqs.(\ref{sr})
one should use as an input the leptonic constants obtained by two-point
QCD sum rules at order $\alpha_s=0$, to be consistent
with the calculation of the correlator
(\ref{corr}) carried out at the same order in $\alpha_s$.
Using $f_D=170 \pm 10 \; MeV$,
$f_{D^*}=220 \pm 24 \; MeV$,
$f_B=150 \pm 20 \; MeV$,
$f_{B^*}=190 \pm 10 \; MeV$,
we obtain:
\bea \gib & = &  20 \pm 4  \label{gbi0}\\
 \gid & = &  9 \pm 1  \; . \label{gdi0} \eea
\noindent
It should be noticed, however, that in a sum rule the
$O(\alpha_s)$ terms appear in the coefficients of the
condensates, where they usually give small corrections, and
in the perturbative contribution, where they are sizeable
in the case of the leptonic constants.
On the other hand, in the determination of
$\gi$ the perturbative contribution (given by the operator $1$ of the expansion
of the correlator in eq.(\ref{corr})
calculated between the vacuum and the pion state)
vanishes at all orders in $\alpha_s$;
this is evident if one considers two-point
functions in the external pion field \cite{eletski}, \cite{grozin}.
For this reason it seems to us that
$\alpha_s$ corrections could be included in the calculation of the leptonic
constants, where they are relevant, whereas  they could
 be safely neglected in all other quantities. Using the results obtained
by  two-point function QCD sum rules (with $O(\alpha_s)$corrections)
$f_B=180 \pm 30 \; MeV$, $f_D=195 \pm 20 \; MeV$
\footnote{ For a review see \cite{paver,dominguez}. }
and $f_{B^*}=213 \pm 34\; MeV$,  $f_{D^*}=258 \pm 26\; MeV$
obtained by a direct calculation in \cite{colangelo1} and
by an analysis of the heavy quark spin symmetry breaking in \cite{neubert1}
we get for $\gi$:
\bea \gib & = &  15 \pm 4  \label{gbi}\\
 \gid & = &  7 \pm 1  \; . \label{gdi} \eea
\noindent

Using the value in eq.(\ref{gdi}) we predict:
$\Gamma (D^{*+} \to D^o \pi^+) = 10 \pm 3 \; KeV$.
Moreover, assuming the dominance of the $B^*$ pole in the form factor
$F_1(q^2)$ for $B \to \pi \ell \nu$, we obtain, using (\ref{gbi}):
$F_1(0)=0.30 \pm 0.07$ in agreement with other QCD sum rules calculations of
this form factor \cite{ball,santorelli}.

Let us discuss an interesting consequence of our result.
In a recent paper \cite{ligeti} it has been suggested that the scaling laws of
the heavy quark effective theory can be used to derive the ratio
$|V_{ub}/V_{cd}|$ from the comparison of the spectra $B \to \pi \ell \nu$
and $D \to \pi \ell \nu$ near the zero recoil point. A crucial ingredient in
this strategy is the ratio
\be
R_{BD} = \sqrt{m_D \over m_B} {f_{B^*} \over f_{D^*} } {\gib \over \gid}
\ee
which is equal to $1$ in the infinite heavy quark mass limit. Our results point
to a value:
\be
R_{BD} = 1.06 \; - \; 1.11
\ee
depending on the role of $\alpha_s$ corrections (the highest value corresponds
to $\alpha_s=0$); this $O(10 \%)$ correction should be taken into account when
determining $V_{ub}$ in this approach.

\resection{The limit $m_Q \to \infty$}

The infinite heavy quark mass limit ($m_B \to \infty$) in eq.(\ref{sr})
can be performed according to the procedure already applied to leptonic
constants \cite{shuryak}, \cite{neubert} and form factors \cite{paver}.
In terms of low energy parameters the quantities in eq.(\ref{sr})
can be written as follows:
\bea
m_B &=& m_b+\omega\nn\\
m_{B^*}-m_B &=& {\cal{O}} \left( {1\over{m_b}}\right)\nn\\
f_B &=& f_{B^*} = {{\hat F}\over{\sqrt {m_b}}} \;
\label{par}
\eea
\noindent $\omega$ represents the binding energy of the meson, which is finite
in the limit $m_b \to \infty$; the scaled leptonic constant $\hat F$ is
independent of $m_b$ (if one neglects a small logarithmic dependence).
$\hat F$ has been computed by QCD sum rules:
for $\omega=0.625$ GeV and the threshold
$y_0 = {s_0 - m_b^2 \over 2 m_b}$ in the range $1.1 - 1.4 \;  GeV$
the result is
 ${\hat F}= 0.30 \pm 0.05 \; GeV^{3/2}$ (at the order $\alpha_s=0$)
\cite{neubert1}
and  ${\hat F}= 0.41 \pm 0.04 \; GeV^{3/2}$
(including radiative corrections) \cite{neubert}.

The sum rule for $g$ in eq.(\ref{ginf}) can be derived from (\ref{sr}) after
having expressed the Borel parameter $M^2$ in terms of the low energy parameter
$E$: $M^2=2 m_b E$. One readily obtains:
\bea
g &=& {{f_{\pi}^2}\over{{\hat F}^2}} e^{\omega /E}\Big\{{1\over{y_0-\omega}}
\Big[ \omega^2 \left( 1 -{\qq\over{3f_{\pi}^2 E}}-{{5 m_1^2}\over{36 E^2}} -
{{m_0^2\qq}\over{72 E^3 f_{\pi}^2}} \right) + \nn\\
&-& 2\omega \left({\qq\over{3f_{\pi}^2}}+{{5 m_1^2}\over{18 E}} +
{{m_0^2\qq}\over{24 E^2 f_{\pi}^2}} \right) -{{5 m_1^2}\over{18}}-{{m_0^2\qq}
\over{12 E f_{\pi}^2}}\Big]+\nn\\
&+& \omega \left( 1 -{\qq\over{3f_{\pi}^2 E}}-{{5 m_1^2}\over{36 E^2}} -
{{m_0^2\qq}\over{72 E^3 f_{\pi}^2}} \right)- {\qq\over{3f_{\pi}^2}}+\nn\\
&-& {{5 m_1^2}\over{18 E}} -{{m_0^2\qq}\over{24 E^2 f_{\pi}^2}} \Big\}
\label{sr2}
\eea

This sum rule must be studied in the region of the external parameter $E$ where
the OPE is assumed to converge and where the contribution of higher resonances
is small ("duality" region).
By imposing the constraint that the various terms of the OPE
in (\ref{sr2}) display a hierarchical structure, according to the dimension, we
get that $E$ should be larger than $1-2 \; GeV$. As for the contribution of
higher resonances, it is controlled in (\ref{sr2}) by the parasitic term. It
can
be shown that this contribution, for values of the parameters in the
range chosen above, is numerically of the same size of the contribution of the
lowest lying states for any value of $E$.
A prediction  for $g$ can be obtained by
studying the stability plateau for (\ref{sr2}), which corresponds to
$E \ge 4 \; GeV$.

We obtain:
\be {\hat F}^2 \; g = 0.035 \pm 0.008 GeV^3 \; .\ee
Therefore at the order $\alpha_s=0$ our result is:
\be g = 0.39 \pm 0.16 \; .\label{ginf0}\ee
As we have discussed already, we could introduce a partial set of
$O(\alpha_s)$ corrections, i.e. those induced by $\hat F$, which should
represent the largest part of such corrections; in this case we get:
\be g = 0.21 \pm 0.06 \; . \label{ginf1}\ee
\noindent The difference between (\ref{ginf0}) and (\ref{ginf1}) reflects
the well known important role of radiative corrections in the determination of
$f_B$ by QCD sum rules in the $m_Q \to \infty$ limit \cite{broad}.
However it should be noticed that the value of $g$ extracted from
(\ref{ginf}) using (\ref{gbi0}) or (\ref{gbi}) is in the range $0.19
- 0.25$, which, if $1/m_Q$ corrections are not anomalously large, points to the
smaller value in (\ref{ginf1}).

\resection{Conclusions}

The strong couplings $g_{P^* P\pi}$ play an important role in heavy meson
phenomenology. They are directly related to the decays of ${P^*}$ into
${P+\pi}$ and they are expected to be important for the semileptonic decays
of ${P}$ into a pion. They are also important inputs in the effective
chiral lagrangians for heavy mesons.
We have estimated the coupling constants $g_{D^* D \pi}$ and
$g_{B^* B \pi}$ in the soft pion limit and for finite heavy quark mass.
Our calculation uses QCD sum rules as applied to the correlator
of a vector and a pseudoscalar current between pion and vacuum. It is
done for b and c states, thus allowing to examine the validity of the
infinite heavy quark mass limit.
We have
considered the contributions of all the operators up to dimension five to the
operator product expansion: it turns out that the higher dimension ones give
small contributions.
The soft pion limit and the single Borel transformation give rise to
parasitic terms which are not exponentially suppressed. They give an essential
correction to the value of the coupling, of the order of the lowest resonance
contribution.
The duality region, where the OPE shows convergence and the higher resonances
give small contributions, is found for values of $M^2$, the Borel parameter,
higher than those usually found in the ordinary sum rules for semileptonic
decays of heavy mesons. This could indicate a slow convergence of the OPE;
nevertheless we have checked that in this region the parasitic terms do not
give a contribution larger than 50 \% to the sum rule.

As we discussed before, the absence of the pure perturbative term in the OPE,
where usually radiative corrections are relevant, seems to indicate that the
largest $\alpha_s$ corrections in the sum rule could be taken into account by
using the radiative corrected values of the leptonic decay constants. In the
case of finite heavy quark mass, the difference between the two choices (i.e.
leptonic decay constants with or without $\alpha_s$ corrections) is within the
uncertainties of our approach, while in the limit of infinite heavy quark mass
there is a factor of two.
Our numerical calculation of the sum rule, in the case when strong radiative
corrections in the values of the leptonic constants are included, leads to
values $g_{B^* B \pi}=15\pm 4 $ and $g_{D^* D \pi}=7\pm 1$. Without
including
these radiative corrections we are lead to $g_{B^* B \pi}=20\pm 4 $ and
$g_{D^* D \pi}=9\pm 1$. The former case leads for the weak vector form factor
between $B$ and $\pi$, $F_{1}(0)$, to a value of $0.30\pm0.07$, in agreement
with previous calculations.
\par
\vspace*{1cm}
\noindent
{\bf Acknowledgements }\\

We would like to thank R. Casalbuoni for discussions during the early stage of
this work.

\end{document}